\documentclass[preprintnumbers,showpacs,amsmath,amssymb,floatfix,prd,onecolumn,superscriptaddress,nofootinbib]{revtex4}
\usepackage{graphicx}
\usepackage{epsfig}
\usepackage{bm}
\usepackage{amsfonts}
\usepackage{epstopdf}
\usepackage{bm}

\begin{document}

\title{ Note on Power-Law Inflation in Noncommutative Space-Time}

\author{Chao-Jun Feng}
\email{fengcj@shnu.edu.cn} 
\affiliation{Shanghai United Center for Astrophysics (SUCA), \\ Shanghai Normal University,
    100 Guilin Road, Shanghai 200234, P.R.China}

\author{Xin-Zhou Li}
\email{kychz@shnu.edu.cn} \affiliation{Shanghai United Center for Astrophysics (SUCA),  \\ Shanghai Normal University,
    100 Guilin Road, Shanghai 200234, P.R.China}

\author{Dao-Jun Liu}
\email{djliu@shnu.edu.cn} 
\affiliation{Shanghai United Center for Astrophysics (SUCA), \\ Shanghai Normal University,
    100 Guilin Road, Shanghai 200234, P.R.China}

\begin{abstract}

In this paper, we propose a new method to calculate the mode functions in the noncommutative power-law inflation model. In this model, all the modes  created when the stringy space-time uncertainty relation is satisfied are  generated inside the Hubble horizon during inflation. It turns out that a linear term describing the noncommutative space-time effect contributes  to the power spectra of the scalar and tensor perturbations. Confronting this model with latest results from \textit{Planck} and BICEP2, we constrain the parameters in this model and we find it is well consistent with observations. 

\end{abstract}

\maketitle


\section{Introduction}

By generating an equation of state with a significant negative pressure before the radiation epoch, inflation  \cite{Guth:1980zm,Linde:1981mu,Albrecht:1982wi} solves a number of cosmological conundrums, such as the horizon, monopole, entropy problems.  After almost thirty-five years of extensive research, inflation is now considered to be a crucial part of the cosmological history of the universe, having affected indelibly its observational features. 
In the simplest inflation model, the early inflating universe is driven by a scalar field called inflaton, which is classically rolling down the hill of its potential. 
Inflation  predicts a nearly scale-invariant primordial scalar perturbation,  which is regarded as  the seed of the large scale structures in present. Furthermore, there could be also a primordial tensor perturbation during the inflation time in principle , which is a signal of testing the primordial gravitational waves.  All of these are essentially from the quantum fluctuations of the scalar field and the curvature of the universe \cite{Feng:2009kb, Feng:2010ya,Cai:2007et}. These small fluctuations are amplified by the nearly exponential expansion, yielding the scalar and tensor primordial power spectra, which can be observed by measuring the Cosmic Microwave Background (CMB), such as the satellite-based Wilkinson Microwave Anisotropy Probe (WMAP) \cite{Hinshaw:2012aka} and  \textit{Planck} \cite{Ade:2013uln} experiments.

Although the observed CMB temperature fluctuations, which are generated by scalar perturbations, already helped us to constrain many inflation models, there are still many compelling models that predict almost the same parameter values, which are consistent with observations. A large number of current CMB experiment efforts now target B-\textit{mode} polarization, which could be only generated by tensor perturbations. Recently, a ground-based  ``Background Imaging of Cosmic Extragalactic Polarization'' experiment has reported their results (BICEP2). They have shown that the observed B-\textit{mode} power spectrum at certain angular scales is well fitted by a lensed-$\Lambda$CDM + tensor theoretical model with tensor-to-scalar ratio $r=0.20^{+0.07}_{-0.05}$, and $r=0$ is disfavoured at $7.0\sigma$ \cite{Ade:2014xna}.

In fact, although there are many inflation models, we still do not known what is the inflaton field.  As a candidate for the theory of everything, string theory should tell us how a successful theory of cosmology can be derived from it. General relativity might break down due to the very high energies during inflation, and corrections from string theory might be needed.  In the non-perturbative string/M theory, any physical process at the very short distance take an uncertainty relation, called  
the stringy space-time uncertainty relation (SSUR):
\begin{equation}\label{equ:ssur}
	\Delta t_p \Delta x_p \geq l_s^2 \,,
\end{equation}
where $l_s$ is the string length scale, and $\Delta t_p = \Delta t$, $\Delta x_p$ are the  uncertainties in the physical time and space coordinates.  It is suggested that the SSUR is a universal property for strings as well as D-branes \cite{yone, Li:1996rp,Yoneya:2000bt}. Unfortunately, we now have no ideas to derive cosmology directly from string/M theory. Brandenberger and Ho \cite{Brandenberger:2002nq} have proposed a variation of space-time noncommutative field theory to realize the stringy space-time uncertainty relation without breaking any of the global symmetries of the homogeneous isotropic universe. If inflation is affected by physics at a scale close to string scale, one expects that space-time uncertainty must leave vestiges in the CMB power spectrum\cite{Huang:2003zp, Tsujikawa:2003gh,Huang:2003hw,Huang:2003fw,Liu:2004qe,Liu:2004xg,Cai:2007bw, Xue:2007bb}.

In this paper, we shall study the power-law inflation in the noncommutative space-time with a different choice of the $\beta_k^\pm$ functions defined below. In this model, it is much  more clear to see the effect of noncommutative space-time and  much easier to deal with the perturbation functions.  A  linear contribution to the power spectra of the scalar and tensor perturbations is given in this model. We also confront this model with latest results from the \textit{Planck} and BICEP2 experiments, and we find this model is well consistent with observations. This paper is organized  as follows. In next section, we will briefly review cosmological perturbation theory in the noncommutative space-time; in Sec.\ref{sec:power} we calculate the power spectra of the power inflation model in the   noncommutative space-time, and compare with observations. In the last section, we will draw our conclusions and give some discussions. And also, in the Appendix.\ref{app:beta}, we presents the detail calculations and discussions on the SSUR algebra and $\beta_k^\pm$ functions.

\section{Perturbations in noncommutative space-time}
 Given a non commutative space-time, the cosmological background will still be described by the Einstein equations since the background fields only depend on the time variable. Assuming a homogeneous and isotropic background, in the following, we will take the Friedmann-Robertson-Walker metric:
\begin{equation}\label{equ:FRW}
ds^2 = -dt^2+a^2(t)dx^2 \,,
\end{equation}
for a spatially flat universe ($K=0$).  Thus the SSUR relation (\ref{equ:ssur}) becomes:
\begin{equation}\label{equ:ssur1}
	\Delta t \Delta x \geq \frac{l_s^2}{ a(t)} \,,
\end{equation}
which is not well defined when $\Delta t$ is large, because the argument $t$ for the scale factor on the r.h.s. changes over time interval $\Delta t$, and it is thus not clear what to use for $a(t)$ in Eq.(\ref{equ:ssur1}). The problem is the same when one uses the conformal time $\eta$ defined by $dt = ad\eta$. Therefore, for later use, a new time coordinate $\tau$ is introduced as
\begin{equation}\label{equ:tau}
	d\tau = a(t) dt \,,
\end{equation}
such that the metric becomes
\begin{equation}\label{equ:metric2}
	ds^2 = -a^{-2}(\tau) d\tau^2 + a^2(\tau)dx^2\,,
\end{equation}
and the SSUR relation is now well defined:
\begin{equation}\label{equ:ssur2}
	\Delta \tau \Delta x \geq l_s^2 \,.
\end{equation}

The action of the perturbations in $3+1$-dimension space-time could be given as the following
\begin{equation}\label{equ:action31}
	S =  \frac{V }{2}\int _{k<k_0}   d\eta d^3k ~ z^{2}(\eta) \bigg( \phi'_{-k}\phi'_{k} - k^2 \phi_{-k} \phi_{k} \bigg)\,,
\end{equation}
where  $V$ is the total spatial coordinate volume and the prime denotes the derivatives with respect to a new time coordinate $\eta $  defined as
\begin{equation}\label{equ:teta1}
	\frac{d\eta}{d\tau} \equiv a^{-2}_{\text{eff}} =  \left( \frac{\beta_k^-}{\beta_k^+}\right)^{1/2} \,.
\end{equation}
Here we have defined 
\begin{equation}\label{equ:beta1}
	\beta^{\pm}_k(\tau) =  a^{\pm2}\left(\tau+\Delta\tau\right) \,, \quad \Delta\tau = l_s^2k \,,
\end{equation}
and $z$ is the so-called ``Mukhanov variable''.  Here we have taken a different form of the $\beta_k^\pm$ functions, which is equivalent to that used in the literatures by the mean of integration, see the Appendix.\ref{app:beta} for detail calculations and discussions. It is now much more easier to deal with the equations of motion  for the field $\phi_k$ :
\begin{equation}\label{equ:eom}
	u_k'' + \left( k^2 - \frac{z''}{z} \right)u_k = 0 \,,
\end{equation}
which could be derived from the action (\ref{equ:action31}). Here the mode function is defined by $u = z \phi_k$.  To calculate the power spectrum of the scalar perturbation, we have $\phi_k = \zeta_k$, $z=z^{(s)} = a \dot\phi/H$ and $u_k^{(s)} = \zeta_k  a \dot \phi /H$, where $\zeta_k$ is the curvature perturbation. While to calculate the power spectrum of the tensor perturbation, we have $\phi_k = h_k/2$, $z=z^{(t)} = a$ and $u_k^{(t)} = a h_k/2$ , where $h_k$ denotes the independent degree of the tensor mode, $h_+$ and $h_\times$. Therefore, the power spectrum of the metric scalar and tensor perturbation are given by
\begin{equation}\label{equ:scalar}
	\mathcal{P}_s = \frac{k^3}{2\pi^2} |\zeta_k|^2  = \frac{k^3}{2\pi^2}\left |\frac{u_k^{(s)}}{z} \right|^2  \,,
\end{equation}
and 
\begin{equation}\label{equ:tensor}
	\mathcal{P}_t = 2\times \frac{k^3}{2\pi^2} |h_k|^2 = \frac{k^3}{\pi^2} \left| \frac{2u_k^{(t)}}{ a} \right|^2 = 8\frac{\dot \phi^2}{H^2}  \left| \frac{u_k^{(t)}}{ u_k^{(s)}} \right|^2 \mathcal{P}_s \,.
\end{equation}

\section{Power-law inflation in noncommutative space-time}\label{sec:power}

\subsection{The model and power spectrum}

The power-law inflation scenario is driven by an exponential potential 
\begin{equation}\label{equ:potential}
	V(\varphi) = V_0 \exp\left(  - \sqrt{\frac{2}{n}} \varphi \right) \,,
\end{equation}
where $V_0$ and $n$ are some constant. For slow-roll inflation, the parameter $n$ should be large enough. Here and after we work in the unit $8\pi G= M_{pl}^{-2}=1$. The corresponding solution of the Friedmann equation is exactly the power-law form
\begin{equation}\label{equ:pinfa}
	a(t) = a_0 t^n = \alpha_0 \tau^{n/(n+1)} \,,
\end{equation}
which is equivalent to the solution when an idea fluid is given with a constant equation of state parameter $w$:
\begin{equation}\label{equ:n}
	n = \frac{2}{3(1+w)}\,, \quad \text{and }\quad \alpha_0 = \big[ a_0(n+1)^n \big]^{1/(n+1)} \,.
\end{equation}
The Hubble parameter $H\equiv \dot a/a$ is given by
\begin{equation}\label{equ:H}
	H = \frac{da}{adt} = \frac{da}{d\tau} =  \frac{n}{n+1}   \alpha_0   \tau^{-1/(n+1)} \,.
\end{equation}
During inflation ( $n$ is large ), we have
\begin{equation}\label{equ:hinf}
	H_{*} \approx \alpha_0\,,
\end{equation}
where $H_*$ denotes the value of $H$ during inflation. By using the definition of $\eta$ and $\beta^\pm$  in Eqs.(\ref{equ:teta1}) and (\ref{equ:beta1}) , we get
$\beta^{\pm}(\tau) =   \alpha_0^{\pm2} \cdot \left(\tau+\Delta\tau\right)^{\pm2n/(n+1)}$ 
and
\begin{equation}\label{equ:etapl}
	 \eta = \alpha_0^{-2}  \int d\tau  \left(\tau+\Delta\tau\right)^{-2n/(n+1)} =  \alpha_0^{-2} \frac{1+n}{1-n} \bigg(\tau+\Delta\tau\bigg)^{(1-n)/(1+n)} \,.
\end{equation}
By using the Friedmann equation $\dot H = -\dot \phi^2/2$, we have $\dot \phi^2/H^2 =2/n$, thus we get $ z^{(t)}  = z^{(s)} \sqrt{ n/2} $. So that the solution of Eq.(\ref{equ:eom}) for the scalar perturbation is the same with that for the tensor perturbation, then we get the tensor-to-scalar ratio 
\begin{equation}
	r \equiv \frac{\mathcal{P}_t }{\mathcal{P}_s}=  \frac{16}{n}  \,.
\end{equation}
The coefficient of the third term in the perturbative equation (\ref{equ:eom}) is then given by 
\begin{equation}
	\frac{ z^{(t) ''}  }{ z^{(t)} }  =  \frac{ z^{(s)''}  }{ z^{(s)} } =  \frac{1}{\eta^{2} } \frac{n(2n-1) }{ (1-n)^2} ( 1- \lambda)^{-2} \bigg ( 1 - \frac{2n}{2n-1}\lambda  \bigg)\,,
\end{equation}
where we have defined
\begin{equation}\label{equ:lambda}
	\lambda =   \frac{\Delta\tau}{\tau + \Delta\tau}  \,,
\end{equation}
which  decreases with time $\tau$. 

In the appendix.\ref{app:beta}, we have shown that all the modes with wave number $k$ are created when  the SSUR is saturated. From Eq.(\ref{equ:k0}), we get the upper bound  of the comoving wave number in the power-law inflation as 
\begin{equation}\label{equ:k01}
	k_0(\tau) = \frac{a_{\text{eff}}}{l_s} = \frac{ a(\tau + \Delta_\tau) }{l_s} = \frac{\alpha_0}{l_s} (\tau+\Delta\tau)^{n/(n+1)} \,,
\end{equation}
which means at time $\tau$, a mode with comoving wave number $k_0$ is created. Then the correspond parameter $\lambda$ during inflation ($n\gg1$) is given by
\begin{equation}\label{equ:approx}
	\lambda_0 \equiv \frac{\Delta\tau}{\tau + \Delta\tau} \bigg|_{k = k_0}= \frac{l_s^2 k_0}{( k_0 l_s/\alpha_0)^{(n+1)/n}} \approx  l_s\alpha_0  \approx \frac{l_s}{H_*^{-1}}
\end{equation}
where $H_*^{-1}$ is the Hubble horizon, see Eq.(\ref{equ:hinf}).  As we discussed earlier, $l_s$ measures the uncertainty between the space and time though SSUR (\ref{equ:ssur}). For example, when one have a determined time $\tau$, there is at least an uncertainty $l_s$ when one measures the distance  $x$. However things are different in the case of an horizon existing. In such case, one can not measure the distance larger than the horizon, e.g. $H_*^{-1}$ since the causality losts and even If  $l_s $ is also larger than the horizon, one can totally lost the prediction of the distance. This is equivalent to the case when $l_s\rightarrow\infty$, which can not be imaged in a real word. Therefore, in the following, we will focus on the case $l_s\ll H_*^{-1}$, or $\lambda_0 \ll1$, one shall see that it is  consistent with observations. Since $\lambda$ decreases with time $\tau$, see Eq.(\ref{equ:lambda}), then  $ \mathcal{O}( |\lambda | ) \sim \mathcal{O}( |\lambda_0 | )  \sim l_s\alpha_0 \ll1$ during the inflation time. Therefore, in the following, we will regard $\lambda$ as a small free parameter in the model, and  keep up to the first order of $\lambda$ in calculations.

At the same time when a mode is created (\ref{equ:k01}),  the wave number cross the comoving Hubble horizon is given by
\begin{equation}\label{equ:cross}
	k_c = a(\tau) H(\tau)  = a'(\tau) a(\tau) =\frac{n}{n+1}  \alpha_0^2  \tau^{2n/(n+1)-1} \,.
\end{equation}
Therefore, during inflation when $n\gg1$,   we have
\begin{equation}\label{equ:k0kc}
	\frac{k_c}{k_0} = \frac{n}{n+1}  l_s \alpha_0   \tau^{-1/(n+1)} (1-\lambda)^{n/(n+1)}  \approx l_s \alpha_0 \approx \lambda \ll 1\,,
\end{equation}
which means all the modes are created inside the horizon in the power-law inflation scenario. Thus, the power spectrum will be calculated at the time when the mode crosses the Hubble horizon ($k=aH$).

The equation of motion(\ref{equ:eom})  could be rewritten as 
\begin{equation}\label{equ:eomapp}
	u_k'' + \left( k^2 - \frac{\nu^2 -1/4}{\eta^2} \right)u_k = 0  \,,
\end{equation}
where
\begin{equation}\label{equ:nu}
	\nu = \frac{3}{2} + \frac{1}{n} + \frac{2}{3}\lambda \,,
\end{equation}
up to the first order of $1/n$ and $\lambda$.  With the initial Bunch-Davies vacuum condition:
\begin{equation}\label{equ:init}
	u_k = \frac{1}{\sqrt{2k}} e^{-ik\eta}\,,
\end{equation}
we get the solution to Eq.(\ref{equ:eomapp}) 
\begin{equation}\label{equ:sol}
	u_k(\eta) = \frac{\sqrt{\pi}}{2} e^{i(\nu+1/2)\pi/2} \sqrt{-\eta} H_\nu^{(1)} (-k\eta) \,,
\end{equation}
where $H_\nu^{(1)}$ is the Hankel's function of the first kind. At the superhorizon scales the solution becomes
\begin{equation}\label{equ:superhorizon}
	u_k(\eta) = 2^{\nu-3/2}  e^{i(\nu-1/2)\pi/2}  \frac{\Gamma(\nu)}{\Gamma(3/2)} \frac{1}{\sqrt{2k}} (-k\eta)^{1/2-\nu} \,.
\end{equation}
Then the power spectrum of the scalar perturbation is given by Eq.(\ref{equ:scalar}) as follows
\begin{equation}\label{equ:scalarsol}
	\mathcal{P}_s = 2^{2\nu-4} n \left[\frac{\Gamma(\nu)}{\Gamma(3/2)} \right]^2 \left(\frac{H}{2\pi}\right)^2 \left(\frac{k}{aH}\right)^{3-2\nu} \bigg|_{k=aH} \approx \frac{n}{8\pi^2} \frac{ H^2 }{M_{pl}^2}\bigg|_{k=aH}\,,
\end{equation}
while the power spectrum of the tensor perturbation is given by Eq.(\ref{equ:tensor}) as follows
\begin{equation}\label{equ:tensorsol}
	\mathcal{P}_t =  2^{2\nu}  \left[\frac{\Gamma(\nu)}{\Gamma(3/2)} \right]^2 \left(\frac{H}{2\pi}\right)^2 \left(\frac{k}{aH}\right)^{3-2\nu} \bigg|_{k=aH} \approx \frac{2}{\pi^2} \frac{ H^2 }{M_{pl}^2}\bigg|_{k=aH}\,
\end{equation}
Therefore the spectrum index of the power spectrum for scalar  $n_s-1 \equiv d\ln \mathcal{P}_s/d\ln k$ and tensor perturbations  $n_t \equiv d\ln \mathcal{P}_t/d\ln k$are  given by
\begin{equation}\label{equ:nst}
	n_s = 1 + 3-2\nu = 1 - \frac{2}{n} - \frac{4}{3}\lambda  \,,\quad  n_t = - \frac{2}{n} - \frac{4}{3}\lambda \,.
\end{equation}
The consistency relation becomes
\begin{equation}\label{equ:cr}
	r = - 8 \left(n_t + \frac{4}{3}\lambda\right)  \,, \quad \text{or} \quad r= -8 \left(n_s-1 + \frac{4}{3}\lambda\right) \,.
\end{equation}
When $\lambda\rightarrow0$, it reduces to the one in the commutative case, i.e. $r = -8n_t$. One shall see that with the help of $\lambda$ term in the above equation, the power-law inflation in noncommutative space-time may be more consistent with observations than that in the commnutative case.

\subsection{Confront the model with observations}
\begin{figure}
\centering
\includegraphics[width=0.4\linewidth]{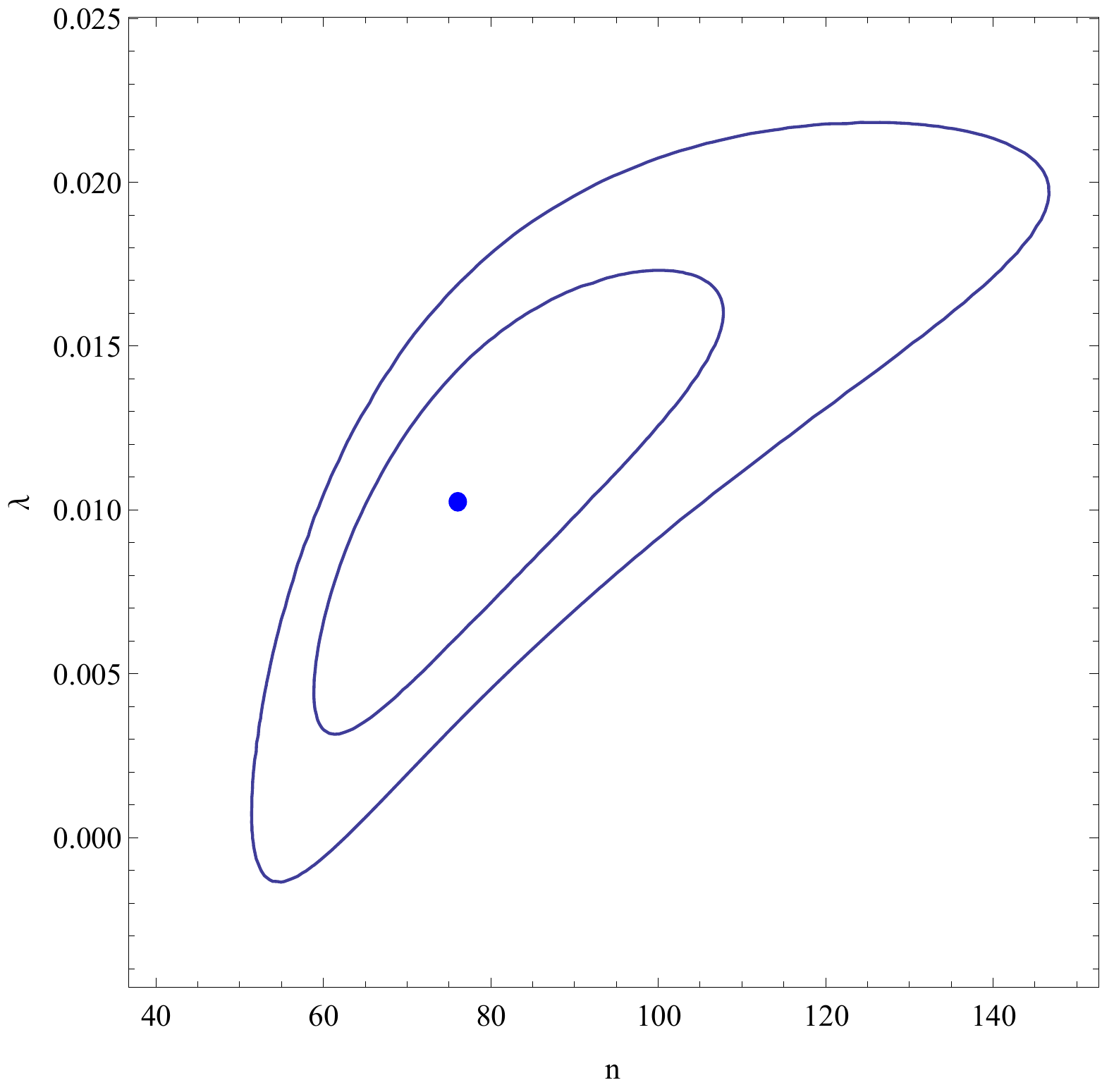}
\quad
\includegraphics[width=0.5\linewidth]{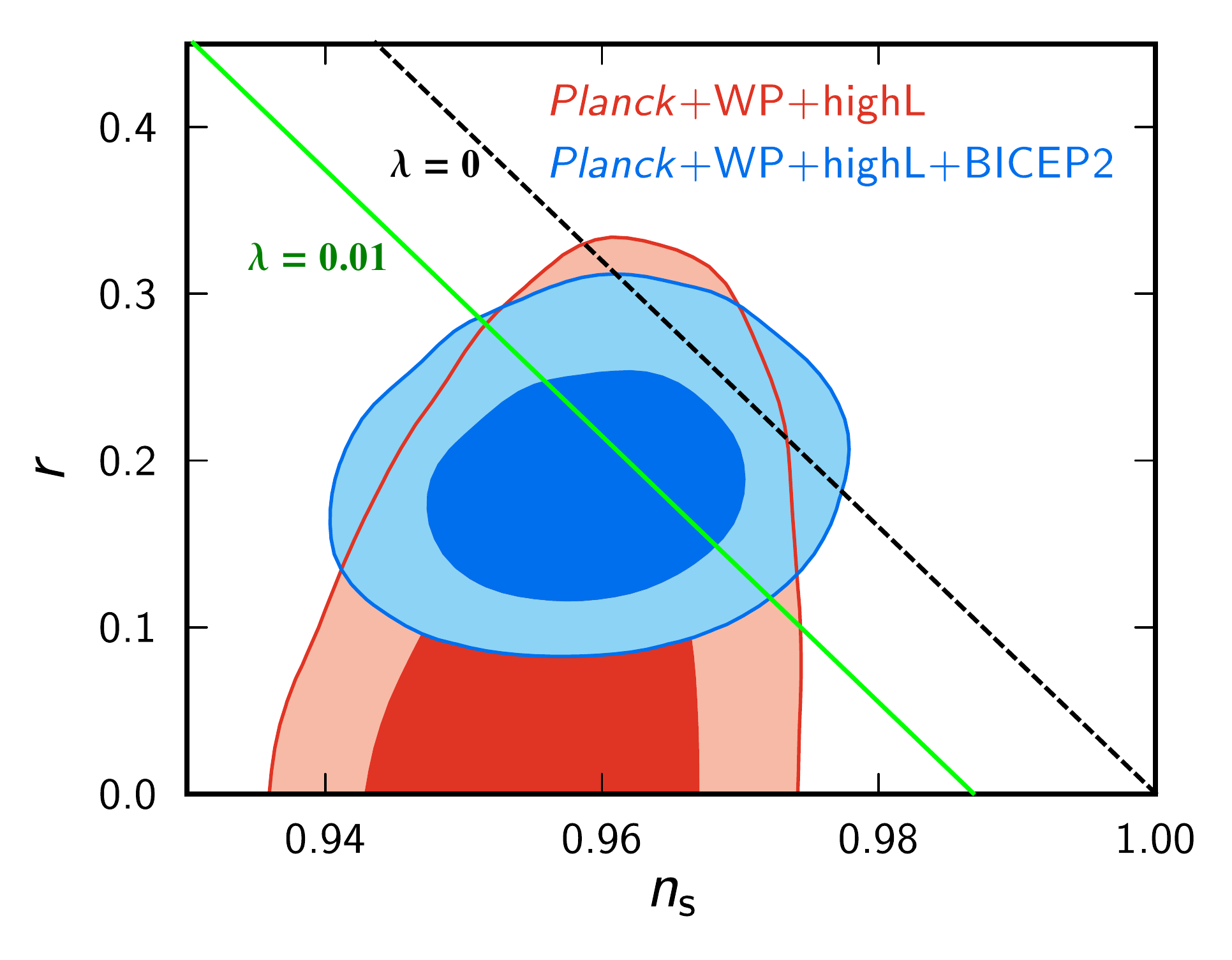}
\caption{Left: Constraints on the values of $n$ and $\lambda$. Two constraint contours are given at $68\%$ and $95\%$ confidence level. The central dot corresponds to  the best-fit point($n=76.1$,$\lambda=0.0102$). Right: the $n_s$-$r$ plane based on  Fig.13 from Ref.\cite{Ade:2014xna}, in which the red contours are simply the Monte Carlo Markov Chains provided with the \textit{Planck} data release, while the blue one is plotted when the BICEP2 data are added. The green solid line corresponds to the noncommutative power-law inflation model (with $\lambda=0.01$) , while the dark dashed line coresponds to the commutative one (with $\lambda=0$) . }
\label{fig:fig1}
\end{figure}

In the following, we will constrain the noncommutative power-law inflation by using the analyse results from data including the $Planck$ CMB temperature likelihood supplemented by the WMAP large scale polarization likelihood (henceforth $Planck$+WP). Other CMB data extending the \textit{Planck} data to higher-$l$, the \textit{Planck} lensing power spectrum, and BAO data are also combined, see Ref.\cite{Ade:2013uln} for details. In Ref.\cite{Ade:2013uln}, the index of scalar power spectrum is given by: $ 0.9583\pm0.0081$(\textit{Planck}+ WP), $ 0.9633\pm0.0072$(\textit{Planck}+WP+ lensing), $ 0.9570\pm0.0075$(\textit{Planck}+WP+highL), $ 0.9607\pm0.0063$(\textit{Planck} +WP+BAO). From the recent reports of BICEP2 experiment, we get the tensor-scalar-ratio as $r=0.20^{+0.07}_{-0.05}$, see Ref.\cite{Ade:2014xna} for details. Also, adopting the data from BICEP2 together with \textit{Planck}  and WMAP  polarization data, Cheng and Huang \cite{Cheng:2014ota} got the constraints of $r=0.23^{+0.05}_{-0.09}$, and  $n_t=0.03^{+0.13}_{-0.11}$. By using these results,  we obtain the constraints on the parameters $n$ and $\lambda$ as 
\begin{equation}\label{equ:constrain}
n=76.14^{+31.65}_{-17.28} \,,\quad \lambda=0.0102^{+0.0058}_{-0.0058}\,, \quad (68\% \text{CL}) \,.
\end{equation}
We plot the contours from $1\sigma$ to $2\sigma$ confidence levels for the parameters,  see Fig.\ref{fig:fig1}, in which the $n_s$-$r$ plane that based on Fig.13 from Ref.\cite{Ade:2014xna} is also presented.  From Fig.\ref{fig:fig1}, one can see that the noncommutative power law inflation with its best fitting parameters is well consistent with observations, while the commutative one ($\lambda=0$) lies outside the $1\sigma$ contour.

\section{Discussion and conclusions}
In this paper, we  suggest to take the first form of the $\beta_k^{\pm}$ functions,  see Eq.(\ref{equ:beta}).  By using this form, it is much more clear to see the effect of noncommutative space-time and much easier to deal with the perturbation functions. A linear contribution to the power spectra of the scalar and tensor perturbations is found in this model. In fact, the second form in Eq.(\ref{equ:beta}) could be also taken by simply redefining $k$ to $-k$, and the results will not be changed. The approximation used in the power-law inflation is $\lambda\ll1$, where $\lambda$ is a free parameter describing the noncommutative effect,see Eq.(\ref{equ:approx}). In other words, all the modes  created when the stringy space-time uncertainty relation is satisfied are  generated inside the Hubble horizon during inflation. It is not  necessarily  to consider the case that all the modes are  generated outside the Hubble horizon, because in this case all the modes have no causality to each other, and then the flat problem in Big Bang theory can not be solved.

After confronting the noncommutative power-law model with the latest results from \textit{Planck} and BICEP2, we constrained the parameter $n$ and $\gamma$, see Fig.\ref{fig}. We conclude that  the model is well consistent with the observations. Using the amplitude value of the power spectrum from \textit{Planck}, $\mathcal{R}_s (k=0.002 Mpc^{-1}) = 2.215\times 10^{-19}$ \cite{Ade:2013zuv}, we can also estimate the value of Hubble parameter during inflation of 
\begin{equation}\label{equ:infhubble}
	\frac{H_*}{M_{pl}} = \pi \sqrt{r \mathcal{R}_s/2 } \approx 4.67 \times 10^{-5} \,,
\end{equation}
where $r=0.20$ was used. Then, by using the fitting value of $\lambda \approx 0.01 $, we estimate the string scale as $l_s \approx 2.14 \times 10^2 l_p \approx 1.7\times 10^{-30}$cm, which is a little smaller than that in Refs.\cite{Huang:2003hw, Liu:2004qe}.

\appendix
\section{The SSUR algebra and $\beta^\pm_k$ functions}\label{app:beta}
In the case of $1+1$ dimension space-time, the SSUR (\ref{equ:ssur}) can be realized by the algebra 
\begin{equation}\label{equ:algbra}
	[\tau, x]_* = il_s^2\,,
\end{equation}
with the $*$ product defined as 
\begin{equation}\label{equ:star}
	(f*g)(\tau, x) = \exp\bigg[ -\frac{i}{2} l_s^2\left(\partial_x\partial_{\tau'} - \partial_\tau\partial_{x'}\right) \bigg] f(\tau, x) g(\tau',x') \bigg|_{\tau'=\tau, x'=x} \,.
\end{equation}
Although this new product introduces higher derivatives of time in the Lagrangian of a field theory, it will not break the unitarily. This because here the field theory we consider below is essentially an effective free theory, while the  fundamental theory is string theory, and it is very common for effective theory to have higher derivative terms.

The free field action for a real scalar field in $1+1$ dimensions is given by
\begin{equation}\label{equ:action11}
	S = \frac{1}{2}\int d\tau dx \bigg[ (\partial_\tau\phi)^\dag * a^2 *( \partial_\tau\phi) -  (\partial_x\phi)^\dag * a^{-2} * (\partial_x\phi) \bigg] \,.
\end{equation}
By expanding the scalar field in Fourier mode, 
\begin{equation}\label{equ:fourier}
	\phi(\tau, x) = V^{1/2} \int _{k<k_0}\frac{dk}{2\pi} \phi_k(\tau) e^{-i kx}\,,
\end{equation}
where $V$ is the total spatial coordinate volume. Since the scalar field is real, we have
\begin{equation}\label{equ:fourier}
	\phi(\tau, x)^\dag = V^{1/2} \int _{k<k_0}\frac{dk}{2\pi} \phi_k^\dag(\tau) e^{i kx} = \phi(\tau, x) = V^{1/2} \int _{k<k_0}\frac{dk}{2\pi} \phi_{-k}(\tau) e^{i kx}\,,
\end{equation}
so that we get $\phi_k^\dag = \phi_{-k}$.  By using the $*$ product (\ref{equ:star}), we have
\begin{eqnarray}
\nonumber
(\partial_\tau\phi)^\dag * a^2 &=&  V^{1/2}  \exp\bigg[ -\frac{i}{2} l_s^2\left(\partial_x\partial_{\tau'} - \partial_\tau\partial_{x'}\right) \bigg]\int _{k<k_0}\frac{dk}{2\pi} \partial_\tau\phi_{-k}(\tau) e^{i kx}a^2(\tau') \bigg|_{\tau'=\tau, x'=x}  \\
\nonumber
&=&  V^{1/2} \int _{k<k_0}\frac{dk}{2\pi} \partial_\tau\phi_{-k}(\tau) e^{i kx} \exp\bigg( \frac{l_s^2k}{2} \partial_{\tau'}  \bigg) a^2(\tau') \bigg|_{\tau'=\tau, x'=x} \\
&=&  V^{1/2} \int _{k<k_0}\frac{dk}{2\pi}\partial_\tau \phi_{-k}(\tau) e^{i kx} a^2\left(\tau+\frac{l_s^2k}{2}\right)  \,,
\end{eqnarray}
and
\begin{eqnarray}
\nonumber
&&(\partial_\tau\phi)^\dag * a^2 * (\partial_\tau \phi) \\ 
\nonumber
&=&  V \exp\bigg[ -\frac{i}{2} l_s^2\left(\partial_x\partial_{\tau'} - \partial_\tau\partial_{x'}\right) \bigg] \int _{k,k'<k_0}\frac{dkdk'}{(2\pi)^2}\partial_\tau \phi_{-k}(\tau) e^{i kx} a^2\left(\tau+\frac{l_s^2k}{2}\right)\partial_{\tau'}\phi_{k'}(\tau') e^{-i k'x'} \bigg|_{\tau'=\tau, x'=x}  \\
\nonumber
&=&  V \int _{k,k'<k_0}\frac{dkdk'}{(2\pi)^2} \exp\bigg[ \frac{l_s^2}{2} \left(k\partial_{\tau'} + k' \partial_\tau\right) \bigg]  \partial_\tau \phi_{-k}(\tau) e^{i kx} a^2\left(\tau+\frac{l_s^2k}{2}\right)\partial_{\tau'}\phi_{k'}(\tau') e^{-i k'x'} \bigg|_{\tau'=\tau, x'=x}  \\
&=&  V \int _{k,k'<k_0}\frac{dkdk'}{(2\pi)^2} \partial_\tau \phi_{-k}\left(\tau+\frac{l_s^2k'}{2}\right)e^{-i (k'-k)x} a^2\left(\tau+l_s^2k\right)\partial_{\tau}\phi_{k'}\left(\tau+\frac{l_s^2k}{2}\right) \,.
\end{eqnarray}
Therefore, the first term of the integration in the action (\ref{equ:action11}) becomes
\begin{eqnarray}
\nonumber
&&\frac{1}{2}\int d\tau dx (\partial_\tau\phi)^\dag * a^2 * (\partial_\tau \phi) \\ 
\nonumber
&=&  \frac{V}{2} \int _{k,k'<k_0}\frac{dkdk'}{(2\pi)^2} d\tau \left[ \partial_\tau \phi_{-k}\left(\tau+\frac{l_s^2k'}{2}\right) a^2\left(\tau+l_s^2k\right)\partial_{\tau'}\phi_{k'}\left(\tau+\frac{l_s^2k}{2}\right)  \right]  \int dx e^{-i (k'-k)x} \\
\nonumber
&=&  \frac{V}{2} \int _{k,k'<k_0}\frac{dkdk'}{(2\pi)^2} d\tau \left[ \partial_\tau \phi_{-k}\left(\tau+\frac{l_s^2k'}{2}\right) a^2\left(\tau+l_s^2k\right)\partial_{\tau'}\phi_{k'}\left(\tau+\frac{l_s^2k}{2}\right)  \right] 2\pi \delta(k'-k)\\
&=&   \frac{V}{2\pi}\int _{k<k_0}\frac{dk}{2} d\tau \left[ \partial_\tau \phi_{-k}\left(\tau+\frac{l_s^2k}{2}\right) a^2\left(\tau+l_s^2k\right)\partial_{\tau}\phi_{k}\left(\tau+\frac{l_s^2k}{2}\right)  \right] \,.
\end{eqnarray}
Let $\tau' = \tau - l_s^2 k/2$ and $k' = k/2$, the above equation becomes
\begin{eqnarray}\label{equ:equiv}
\nonumber
\frac{1}{2}\int d\tau dx (\partial_\tau\phi)^\dag * a^2 * (\partial_\tau \phi) 
\nonumber
&=& \frac{V}{2\pi}\int _{k<k_0}dk d\tau \left[ \partial_\tau \phi_{-k}\left(\tau\right) a^2\left(\tau+l_s^2k\right)\partial_{\tau}\phi_{k}\left(\tau\right)  \right] \\
\nonumber
&\overset{k \rightarrow-k}{=}& \frac{V}{2\pi}\int _{k<k_0}dk d\tau \left[ \partial_\tau \phi_{k}\left(\tau\right) a^2\left(\tau-l_s^2k\right)\partial_{\tau}\phi_{-k}\left(\tau\right)  \right] \\
&=& \frac{V}{2\pi}\int _{k<k_0}dk d\tau \left\{ \partial_\tau \phi_{k}\left(\tau\right) \frac{1}{2}\bigg[a^2\left(\tau-l_s^2k\right) + a^2\left(\tau+l_s^2k\right) \bigg]\partial_{\tau}\phi_{-k}\left(\tau\right)  \right\} \,,
\end{eqnarray}
where we have used the invariant measure 
\begin{equation}
	\int_{-\infty}^\infty dk =- \int_{\infty}^{-\infty} dk' = \int_{-\infty}^\infty dk'
\end{equation}
after $k \rightarrow k'=-k$.  By using the same procedure, one could get the second term of the integration in the action (\ref{equ:action11}). Finally, we get the action as
\begin{equation}\label{equ:action112}
	S =  \frac{V }{2\pi}\int _{k<k_0}  d\tau dk  \bigg[ \beta^+_k \partial_\tau\phi_{-k}\left(\tau\right)\partial_{\tau}\phi_{k}\left(\tau\right) - k^2  \beta^-_k \phi_{-k}\left(\tau\right) \phi_{k}\left(\tau\right) \bigg]\,.
\end{equation}
Here, it should be noticed that the $\beta^\pm_k$ functions could be any taken any of the following form
\begin{equation}\label{equ:beta}
	\beta^{\pm}_k= a^{\pm2}\left(\tau+\Delta\tau\right) \,, \quad 
	\beta^{\pm}_k= a^{\pm2}\left(\tau-\Delta\tau\right)\,, \quad 
	\text{or}\quad
	\beta^{\pm}_k= \frac{1}{2} \bigg[ a^{\pm2}\left(\tau-\Delta\tau\right) + a^{\pm2}\left(\tau+\Delta\tau\right)\bigg]\,,
\end{equation}
since they are equivalent   by the mean of integration, see Eq.(\ref{equ:equiv}). Here $\Delta\tau = l_s^2k$ denotes the uncertainty in time $\tau$. We believe that once we clearly get the exact solution to the perturbation function, there is no difference to take any forms of the $\beta^\pm_k$ function. However, it is hard to obtain the exact solution, then we need to do some approximation, which is depends on the specific form of $\beta^\pm_k$. So far as we known, the third form of $\beta^\pm_k$ in Eq.(\ref{equ:beta}) is often used in the literatures. However, in our paper, we  suggest to take the first form, which seems much more easier to deal with the perturbation functions and solutions, and which seems to be more consistent with observations.

The reason to impose an upper bound on the comoving momentum $k$ at $k_0$ in Eq.(\ref{equ:fourier}) is as follows. A fluctuation mode with wave number $k$ will exist when the SSUR is satisfied. In other words, the mode will be created when the SSUR is saturated. According to Eq.(\ref{equ:action112}), the energy defined with respect to $\tau$ for a given mode $k$ is 
\begin{equation}\label{equ:energy}
	E_k = \frac{ k } {a^{2}_{\text{eff}} }\,,  \quad \text{with} \quad a_{\text{eff}}^2 \equiv \left( \frac{\beta_k^+}{\beta_k^-}\right)^{1/2}\,.
\end{equation}
Buy using the approximation $\Delta x \sim 1/k$, $\Delta \tau \sim 1/E_k$ and the SSUR, we get
\begin{equation}
	\Delta\tau \Delta x \sim \left( \frac{a_{\text{eff}}}{k} \right)^2\geq l_s^2 \,,
\end{equation}
and then the upper bound of the wave number is 
\begin{equation}\label{equ:k0}
	k\leq k_0(\tau) \equiv  \frac{a_{\text{eff}}}{l_s} \,.
\end{equation}

To calculate the power spectrum, it is convenient to rewrite the action in the form
\begin{equation}\label{equ:action113}
	S =  \frac{V }{2\pi}\int _{k<k_0}   d \eta dk ~ y_k^2(\tilde\eta) \bigg( \phi'_{-k}\phi'_{k} - k^2 \phi_{-k} \phi_{k} \bigg)
\end{equation}
where  the prime denotes the derivatives with respect to a new time coordinate $\eta $  defined as
\begin{equation}\label{equ:teta}
	\frac{d \eta}{d\tau} \equiv a^{-2}_{\text{eff}} =  \left( \frac{\beta_k^-}{\beta_k^+}\right)^{1/2} \,, \quad \text{and} \quad y_k = (\beta_k^-\beta_k^+)^{1/4} \,.
\end{equation}
When the string length scale $l_s\rightarrow 0$, the action (\ref{equ:action113}) becomes the one in commutative case:
\begin{equation}\label{equ:actioncomm}
	S =  \frac{V }{2\pi}\int _{k<k_0}   d\tilde\eta dk ~  \bigg( \phi'_{-k}\phi'_{k} - k^2 \phi_{-k} \phi_{k} \bigg)
\end{equation}
where $d\tilde\eta = dt/a$ is the conformal time. Therefore, the previous section motivates a model to incorporate the SSUR for any space-time dimension:
\begin{equation}\label{equ:action1d}
	S =  \frac{V }{(2\pi)^d}\int _{k<k_0}   d \eta d^dk ~ z_k^{d-1}(\tilde\eta) \bigg( \phi'_{-k}\phi'_{k} - k^2 \phi_{-k} \phi_{k} \bigg)\,,
\end{equation}
where $d\eta = a_{\text{eff}}^{-2} d\tau$ 
and
\begin{equation}\label{equ:zkd}
	z_k^{d-1}(\eta) = z^{d-1} y_k^2(\eta)   \,.
\end{equation}
Here $z_k$ is some smeared version of $z$ or $a$ over a range of time of characteristic scale $\Delta \tau = \l_s^2k$. It is supposed that the only difference between the $d+1$ dimension action and the $1+1$-dimension one is the measure $z^{d-1}$ for additional $(d-1)$ dimensions. In the case of gravitational waves, the function $z_k$ is denoted as $a_k$, with $a_k$ constructed from the scale factor $a$ in the same way as $z_k$ is obtained from $z$. Also, it is clearly that $y_k^2=1$ when we choose the first or the second form of $\beta_k^\pm$ in Eq.(\ref{equ:beta}).

\acknowledgments

This work is supported by National Science Foundation of China grant Nos.~11105091 and~11047138, ``Chen Guang" project supported by Shanghai Municipal Education Commission and Shanghai Education Development Foundation Grant No. 12CG51, National Education Foundation of China grant  No.~2009312711004, Shanghai Natural Science Foundation, China grant No.~10ZR1422000, Key Project of Chinese Ministry of Education grant, No.~211059,  and  Shanghai Special Education Foundation, No.~ssd10004, the Program of Shanghai Normal University (DXL124), and Shanghai Commission of Science and technology under Grant No.~12ZR1421700.

\end{document}